\documentclass{emulateapj}
\usepackage{apjfonts}
\usepackage[pagebackref=false,colorlinks=true,citecolor=blue,linkcolor=blue,breaklinks=true,bookmarks=true]{hyperref}

\usepackage{amsmath}
\usepackage{mathtools}   
\usepackage{booktabs}

\newcommand{\um}{$\mu$m}
\newcommand{\kms}{{km~s$^{-1}$}}

\newcommand{\msun}{${M}_{\odot}$}

\newcommand{\beq}{\begin{equation}}
\newcommand{\eeq}{\end{equation}}

\newcommand{\hi}{\ion{H}{1}}
\newcommand{\aco}{\alpha_{\rm CO}}
\newcommand{\fpm}{f^{\prime}_{\rm mol}}

\newcommand{\lco}{$L^{\prime}_{\rm CO}$}

\newcommand{\cohh}{CO$-$H$_2$}
\newcommand{\mstar}{M_{\star}}

\newcommand{\mL}{{\mathcal{L}}}

\newcommand{\herschel}{{\it Herschel}}

\newcommand{\iras}{{\it IRAS}}

\newcommand{\emcee}{{\tt emcee}}

\begin{document}

\title{The Evolution of Molecular Gas Fraction Traced by the CO Tully-Fisher Relation}

\author{Jacob W. Isbell\altaffilmark{1}$^,$\altaffilmark{2}, Rui Xue\altaffilmark{1}, and Hai Fu\altaffilmark{1}}

\altaffiltext{1}{Department of Physics \& Astronomy, The University of Iowa, 203 Van Allen Hall, Iowa City, IA 52242, USA}
\altaffiltext{2}{Max-Planck-Institut f{\"u}r Astronomie, K{\"o}nigstuhl 17, Heidelberg D-69117, Germany}

\begin{abstract}
Carbon monoxide (CO) observations show a luminosity$-$line-width correlation that evolves with redshift. We present a method to use CO measurements alone to infer the molecular gas fraction ($f_{\rm mol}$) and constrain the CO$-$H$_2$ conversion factor ($\alpha_{\rm CO}$). We compile from the literature spatially integrated low-$J$ CO observations of six galaxy populations, including a total of 449 galaxies between $0.01 \leq z \leq 3.26$. The CO data of each population provide an estimate of the $\aco$-normalized mean molecular gas fraction ($f_{\rm mol}/\alpha_{\rm CO}$). The redshift evolution of the luminosity$-$line-width correlation thus indicates an evolution of $f_{\rm mol}/\alpha_{\rm CO}$. We use a Bayesian-based Monte-Carlo Markov Chain sampler to derive the posterior probability distribution functions of $f_{\rm mol}/\alpha_{\rm CO}$ for these galaxy populations, accounting for random inclination angles and measurement errors in the likelihood function. We find that the molecular gas fraction evolves rapidly with redshift, $f_{\rm mol} \propto (1+z)^\beta$ with $\beta \simeq 2$, for both normal star-forming and starburst galaxies. Furthermore, the evolution trend agrees well with that inferred from the Kennicutt-Schmidt relation and the star-forming main sequence. Finally, at $z < 0.1$ normal star-forming galaxies require a $\sim5\times$ larger $\alpha_{\rm CO}$ than starburst galaxies to match their molecular gas fractions, but at $z > 1$ both star-forming types exhibit sub-Galactic $\alpha_{\rm CO}$ values and normal star-forming galaxies appear more gas-rich than starbursts. Future applications of this method include calibrating Tully-Fisher relations without inclination correction and inferring the evolution of the atomic gas fraction with H\,{\sc i} observations.
\end{abstract}


\section{Introduction}

In the Milky Way, star formation occurs in giant molecular clouds \citep[GMCs; e.g.,][]{McKee07}, which are cold, dense, and self-gravitating structures with a typical spatial extent of $\sim$50\,pc \citep[e.g.,][]{Blitz:1993aa}. As the apparent fuel of star formation, molecular gas plays a key role in galaxy evolution \citep[e.g.,][]{Kennicutt12}. Even though the sub-pc-scale physics of star formation is still not well understood, empirically the star formation rate (SFR) surface density is found to tightly correlate with the molecular gas mass ($M_{\rm mol}$) surface density on kpc scales, following the Kennicutt-Schmidt relation \citep[e.g.,][]{Wong:2002aa,Bigiel:2008aa}.
As a result, the fraction of molecular gas within a galaxy's total baryonic mass ($f_{\rm mol} \equiv M_{\rm mol}/M_{\rm bary}$) determines its specific SFR (sSFR = SFR/$M_\star$). In fact, the redshift evolution of the star-forming main sequence \citep[SFR\,$\propto M_{\star}$; e.g.,][]{Noeske07a} is interpreted as the evolution of the molecular gas fraction \citep[e.g.,][]{Bouche10}. Being able to measure $f_{\rm mol}$ directly as a function of redshift is thus crucial for our understanding of the evolution of star-forming galaxies (SFGs). 

The brightest molecular lines from galaxies are emitted by Carbon Monoxide ($^{12}$C$^{16}$O, hereafter CO) because it is the second most abundant molecule and its low-level rotational transitions are easily excitable in GMC-like environments. Therefore, many previous studies of the molecular gas content of galaxies have used the CO line luminosity to estimate the total molecular gas mass, assuming the two are related by a \cohh\ conversion factor \citep[see][for a review]{Bolatto13}.
Combined with stellar mass estimates from stellar population synthesis modeling of the optical-to-IR spectral energy distributions (SEDs), the molecular gas fraction can be estimated for individual galaxies at significant redshifts \citep[e.g.,][]{Tacconi18}, where the atomic/ionized gas contribution is likely negligible due to higher molecular-to-atomic gas ratios \citep[e.g.,][]{Obreschkow:2009aa,Lagos:2011ab}.
Using this approach, $f_{\rm mol}$ is found to increase with redshift and correlate with sSFR -- albeit with large scatters \citep[e.g.,][]{Tacconi10,Tacconi13} -- as is expected from the Kennicutt-Schmidt relation. However, stellar mass estimates from population synthesis models are known to suffer from a number of systematic degeneracies. In particular, different assumed star-formation histories alone can lead to stellar mass estimates that differ by almost an order-of-magnitude \citep[e.g,][]{Michalowski12}. To avoid such uncertainties, here we propose to estimate the molecular gas fraction using the Tully-Fisher relation. 

The original Tully-Fisher relation \citep[TFR;][]{Tully77} is an empirical correlation between the $B$-band absolute magnitude and the \hi\ 21-cm line width of local disk galaxies. In the past four decades, multiple flavors of the TFR have been established by using luminosities at longer wavelengths \citep[e.g.,][]{Tully12,Sorce13,Tiley:2016aa} and/or by adopting kinematics traced by other emission lines, e.g., [O\,{\sc ii}]$\lambda$3776 \citep[][]{Chiu07} and CO \citep[][]{Schoniger94,Ho07b,Davis11}. In addition, the relation is also expanded to samples at significant redshifts \citep[e.g.,][]{Chiu07,Cresci:2009aa,Ubler17,Turner17}. Given the almost constant mass-to-light ratio in rest-frame near-IR wavelengths, luminosity-based TFR can be converted to mass-based TFRs \citep{McGaugh:2012aa,Zaritsky14a,Tiley:2016aa,Topal18}, providing a method to estimate galaxy masses from simply a line profile.

Here we present a method to measure the evolution of molecular gas fraction in massive galaxies between $0 < z < 3$ using the CO TFR. This {\it Letter} is organized as follows: in \S\,\ref{sec:data} we describe the compilation of galaxy-integrated CO line widths and luminosities, in \S\,\ref{sec:analysis} we introduce the Bayesian-based statistical method to infer the molecular gas fraction from CO data and the TFR, and in \S\,\ref{sec:result} we present our results and discuss the systematic uncertainties. We close with a summary and discuss future applications of this method in \S\,\ref{sec:summary}. 
Throughout we adopt the $\Lambda$CDM cosmology with $\Omega_{\rm{m}} = 0.3, \Omega_{\Lambda} = 0.7$, and $H_0 = 70~\rm{km}~\rm{s}^{-1}~\rm{Mpc}^{-1}$.

\section{Data Compilation} \label{sec:data}

\begin{deluxetable*}{llccccccc}
\tablewidth{0pt}
\tablecaption{Compilation of CO Measurements and Best-Fit Model Parameters
\label{tab:cats}}
\tablehead{
\colhead{Reference\tablenotemark{a}} & 
\colhead{Telescope} & 
\colhead{Redshift Range} &
\colhead{N Sources\tablenotemark{b}} & 
\colhead{CO $J_{\rm up}$} & 
\colhead{$\log f^{\prime}_{\rm mol}$ \tablenotemark{c}} &
\colhead{$\sigma$ \tablenotemark{d}} & 
\colhead{$f_{\rm mol}$ \tablenotemark{e}}\\
\colhead{} & \colhead{} & \colhead{} & \colhead{} & \colhead{} & 
\colhead{$\log(L_l\,M_\odot^{-1})$} & \colhead{[dex]} & \colhead{}
}
\startdata
\toprule
\multicolumn{2}{l}{\bf{Local Star-forming Galaxies}}& $0.03 - 0.05$ & {\bf 214} & 1  & $-1.85_{-0.04}^{+0.05}$ & $0.46_{-0.04}^{+0.04}$ & $1.4_{-0.2}^{+0.1}$\%\\
\citet{Saintonge17} & IRAM 30m & $0.03 - 0.05$ & 214 & 1 & & & & \\
\vspace{-7pt} \\
\multicolumn{2}{l}{\bf{Local ULIRG}}& $0.01 - 0.09$ & {\bf 68} & 1  & $-1.18_{-0.10}^{+0.10}$ & $0.62_{-0.07}^{+0.08}$ & $6.7_{-1.7}^{+1.3}$\%\\
\citet{Yamashita17} & NRO 45m & $0.01 - 0.09$ & 68 & 1 & & & & \\
\vspace{-7pt} \\
\multicolumn{2}{l}{\bf{Intermediate Redshift Star-forming Galaxies}}& $0.03 - 0.33$ & {\bf 49} & 1  & $-0.90_{-0.09}^{+0.09}$ & $0.48_{-0.07}^{+0.08}$ & $12.6_{-2.4}^{+2.9}$\%\\
\citet{Villanueva17} & ALMA & $0.03 - 0.33$ & 49 & 1 & & & & \\
\vspace{-7pt} \\
\multicolumn{2}{l}{\bf{Intermediate Redshift ULIRGs}}& $0.22 - 0.91$ & {\bf 36} & 1,2,3  & $-0.52_{-0.12}^{+0.12}$ & $0.60_{-0.08}^{+0.10}$ & $30.3_{-9.7}^{+7.4}$\%\\
\citet{Combes11,Combes13} & IRAM 30m & $0.61 - 0.91$ & 28 & 2,3 & & & & \\
\citet{Magdis14} & IRAM 30m & $0.22 - 0.44$ & 8 & 1,2,3 & & & & \\
\vspace{-7pt} \\
\multicolumn{2}{l}{\bf{High Redshift Star-forming Galaxies}}& $1.00 - 2.43$ & {\bf 51} & 3  & $-0.36_{-0.10}^{+0.11}$ & $0.57_{-0.08}^{+0.09}$ & $44.1_{-12.2}^{+9.4}$\%\\
\citet{Tacconi13} & PdBI & $1.00 - 2.43$ & 51 & 3 & & & & \\
\vspace{-7pt} \\
\multicolumn{2}{l}{\bf{High Redshift Submillimeter Galaxies}}& $1.19 - 3.26$ & {\bf 31} & 1,2,3  & $-0.54_{-0.09}^{+0.09}$ & $0.36_{-0.06}^{+0.08}$ & $28.7_{-6.7}^{+5.4}$\%\\
\citet{Bothwell13} & PdBI & $1.19 - 3.10$ & 19 & 2,3 & & & & \\
\citet{Harris12} & GBT 100m & $2.19 - 3.26$ & 12 & 1 & & & & 
\enddata
\tablecomments{
$^{\rm a}$ The six galaxy populations in our study are highlighted in bold font, followed by the  included references. 
$^{\rm b}$ Bold face values indicate the total number of sources in each population.
$^{\rm c}$ The best-fit $\aco$-normalized molecular gas fraction $\log \fpm=\log (f_{\rm mol}/\alpha_{\rm CO})$ (see \S\,\ref{sec:analysis}).
$^{\rm d}$ The best-fit dispersion parameter $\sigma$ in dex, which captures the observational errors and the intrinsic scatter in the baryonic TFR (see \S\,\ref{sec:analysis} and~\S\,\ref{sec:sysunc}). 
$^{\rm e}$ The molecular gas fraction calculated from $\log \fpm$ assuming a \cohh\ conversion factor of $\alpha_{\rm CO} = 1.0~M_{\odot}/ L_{l}$.
}
\end{deluxetable*}

\begin{figure}
\includegraphics[width=0.5\textwidth]{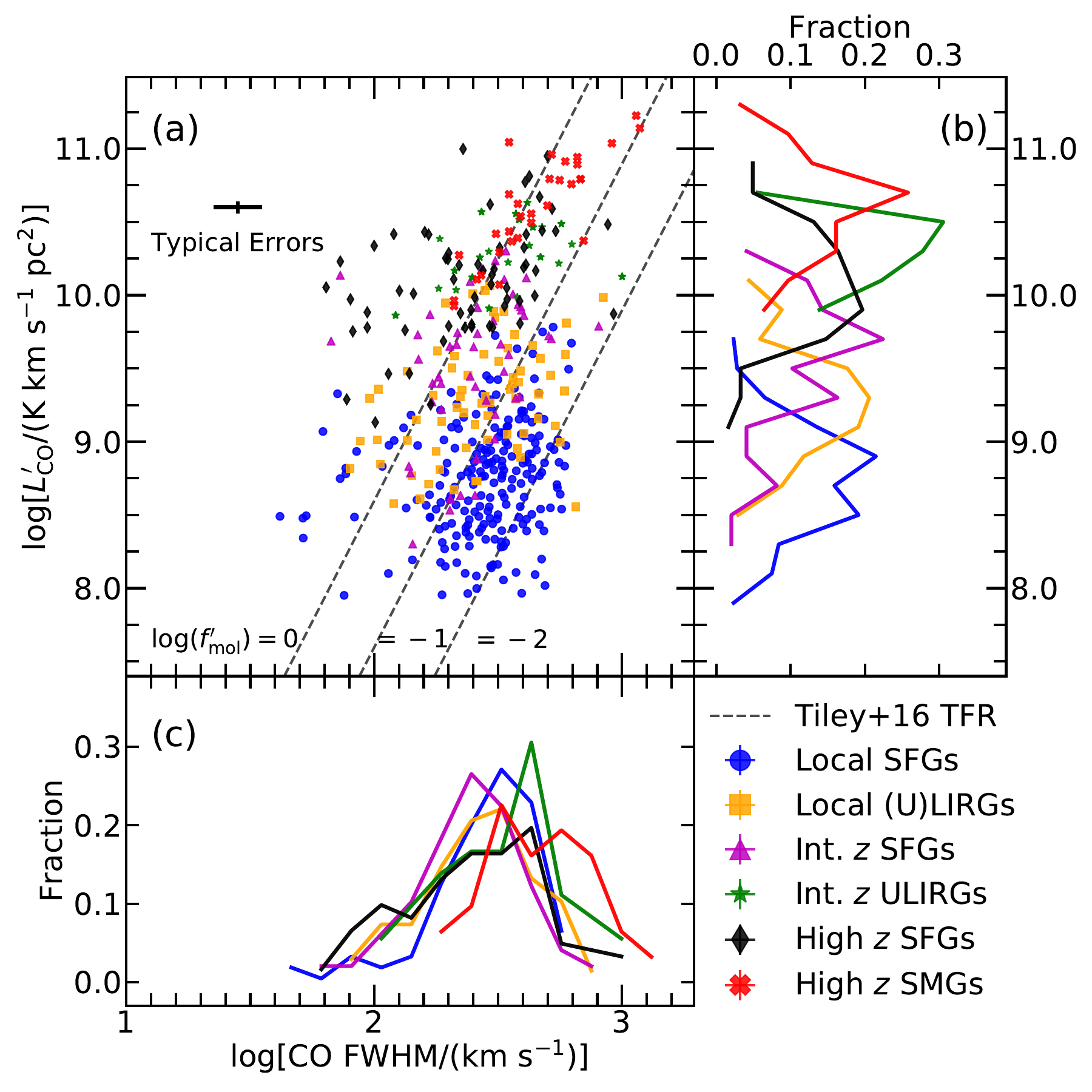}
\caption{Compiled CO measurements of the six galaxy populations in Table\,\ref{tab:cats}. ($a$) CO luminosity (\lco) vs. line FWHM ($w$). Data points are color-coded by their population, as described by the legend. Typical errors are 25\% (0.11 dex) in $w$ and 15\% (0.07 dex) in \lco. The dashed lines show a range of $\fpm \equiv f_{\rm mol} / \alpha_{\rm CO}$ values used to convert line-width-inferred masses to \lco\ (refer to Eqs.\ref{eq:gco} and \ref{eq:tfr}). ($b$) Distributions in line luminosity, plotted in units of fraction per dex for 0.2\,dex bins. ($c$) Distributions in line width, plotted in units of fraction per dex for 0.12\,dex bins.}
\label{fig:lcow50}
\vskip 0.1in
\end{figure}

We have compiled spatially-integrated low-$J$ ($J_{\rm up} \leq 3$) CO measurements of 449 galaxies between $0.01 < z < 3.26$ from the literature. Low-$J$ transitions are preferred over high-$J$ transitions because they have: (1) less extreme excitation conditions, (2) lower scatter in the CO spectral line energy distribution \citep[e.g.,][]{Greve:2014aa}, and (3) more spatially extended distribution \citep[e.g.,][]{Ivison:2011aa,Saito:2017aa}. To carry out our analysis in \S~\ref{sec:analysis}, we only need three observables -- the redshift, the line flux, and the line width. 

The redshift and the line flux coupled with a cosmology give the CO line luminosity (\lco), which is normally defined as the velocity-area-integrated CO brightness temperature \citep{Solomon97}:
\begin{equation}
L^{\prime}_{\rm CO} = 3.25\times10^7~S_{\rm CO} \Delta V~\nu_{\rm obs}^{-2}~D_L^2~(1+z)^{-3} 
\end{equation}
where $S_{\rm CO} \Delta V$ is the integrated line flux in Jy~\kms, $\nu_{\rm obs}$ is the observed frequency in GHz, $D_L$ is the luminosity distance in Mpc. The resulting $L'_{\rm CO}$ is in units of $L_l \equiv {\rm K~\rm km~s}^{-1}~{\rm pc}^2$. To homogenize the different cosmologies adopted in the various references, we have converted the reported \lco\ to our adopted cosmology. Additionally, we convert \lco\ from higher $J$ transitions to the equivalent CO\,($1\rightarrow0$) luminosity using the observed mean correction factor: $R_{J1} \equiv L^{\prime}_{{\rm CO}~J\rightarrow J-1}/L^{\prime}_{{\rm CO}~1\rightarrow0}$. For (ultra-)luminous infrared galaxies (U)LIRGs and submillimeter-bright galaxies (SMGs), we adopt the mean SMG values of $R_{21} = 0.85$ and $R_{31} = 0.66$; and for the normal SFGs, we adopt $R_{21} = 0.9$ and $R_{31} = 0.6$, which are the mean values from high-redshift color-selected SFGs \citep[see Table\,2 in][]{Carilli:2013aa}.

We adopt the full-width at half-maximum (FWHM, hereafter $w$ in \kms) values reported in each reference to characterize CO line widths.
The spectral resolutions are generally much smaller than the line widths, so instrumental broadening has a negligible effect. 
We note that these surveys measure $w$ in various ways, and the values were often estimated from best-fit parametrized models to the actual spectra.
However, as shown by previous studies \citep[e.g.,][]{Bothwell13a, Magdis14,Tiley:2016aa}, there are no systematic offsets among the $w$-values derived from different techniques, because the parameterized models must represent the observed line profiles reasonably well.

The compiled references are listed in Table\,\ref{tab:cats}. We have grouped them into six populations based on their redshift range and source selection criteria. Our compilation is not intended to be complete; instead, we have selected the references which include relatively large numbers of objects ($N \geq 8$) in their corresponding category to minimize inhomogeneity in the data set. There are three populations of normal SFGs and three populations of starburst galaxies: 
\begin{enumerate}

    \item \textit{Local SFGs}: We include CO\,($1\rightarrow0)$ detections of 214 galaxies between $0.025 < z < 0.05$ from the COLD\,GASS survey \citep{Saintonge11,Saintonge17}. This survey used the Institut de Radioastronomie Millim\'etrique (IRAM) 30\,m telescope and targeted 366 stellar-mass-selected galaxies with  $10^{10} M_{\odot} < M_{\star} < 10^{11.5} M_{\odot}$. We deliberately exclude the 166 galaxies in the COLD GASS-low sample ($M_{\star} < 10^{10} M_{\odot}$ and $z < 0.02$), because these galaxies are not massive enough to be comparable with the samples at higher redshifts. The CO line widths were measured using the method of \citet{Springob:2005aa}, which fits a linear slope to each side of the line profile and takes the width at half maximum of these fits.
    
    \item \textit{Local (U)LIRGS}: We include CO($1\rightarrow0$) detections of 68 galaxies in 56 (U)LIRGs with $L_{\rm IR} > 10^{11} L_{\odot}$ between $0.01 <z <0.09$ from \citet{Yamashita17}. The reported CO line widths were measured directly from the emission profiles. These galaxies were observed using the Nobeyama Radio Observatory (NRO) 45\,m telescope and were selected from the Great Observatories All-sky LIRG Survey \citep[GOALS;][]{Armus09}. 

    \item \textit{Intermediate Redshift SFGs}: We include 49 CO\,($1\rightarrow0$) Atacama Large Millimeter Array (ALMA) detections of galaxies between $0.03<z<0.33$ from \citet{Villanueva17}. The galaxies were selected from the \herschel-ATLAS survey to have $>3\sigma$ detections in both the 160\,\um\ and 250\,\um\ bands. The reported line widths are measured from best-fit single-Gaussian profiles. 
            
    \item \textit{Intermediate Redshift ULIRGs}: We include IRAM-30\,m CO detections of 28 galaxies between $0.61<z<0.91$ from \citet{Combes11,Combes13} and 8 galaxies between $0.22<z<0.44$ from \citet{Magdis14}. The galaxies in \citet{Combes11,Combes13} and \citet{Magdis14} are ULIRGs with $L_{\rm IR} > 10^{12.45} L_{\odot}$ and detected at 60$\mu$m (\iras) and 250$\mu$m (\herschel-SPIRE) respectively.  Both samples obtain $w$ from the best-fit single-Gaussian models to the line profiles.
     
    \item \textit{High Redshift SFGs}: We include CO\,($3\rightarrow2$) detections of 51 main-sequence galaxies between $1.00<z<2.43$ from the PHIBBS survey \citep{Tacconi13}. These galaxies have $\mstar > 2.5\times10^{10}~M_{\odot}$ and SFR $> 30~M_{\odot}\,\text{yr}^{-1}$. The authors report the characteristic circular velocity ($v_{\rm c}$) estimated from either the line FWHM for unresolved sources or the inclination-angle-corrected velocity gradient for resolved sources. Because the line FWHMs, the velocity gradients, and the inclination angles are not listed in their tables, we use the isotropic virial estimate adopted by the authors for all their galaxies to convert $v_{\rm c}$ to $w$: $w/v_{\rm c} = \sqrt{(8 \ln 2)/3} = 1.36$. 
    
    \item \textit{High Redshift SMGs}: We include 19 CO detections of SMGs ($S_{850} > 1$\,mJy) between $1.19 < z < 3.10$ from \citet{Bothwell13a}.
Line widths are from the intensity-weighted second moment of each CO spectrum, converted to the equivalent Gaussian $w$.
The authors argue that this method is better for low signal-to-noise spectra, where Gaussian fits may fail to achieve sensible results.
We also include CO\,($1\rightarrow0$) detections of 12 \herschel-selected bright SMGs using the 100-m Green Bank Telescope (GBT) between $2.19<z<3.26$ from \citet{Harris12}.  The reported line widths come from the best-fitted single-Gaussian profiles. Because these \herschel\ galaxies are gravitationally lensed, we correct the observed \lco\ using the magnification factors from the lens models of \citet{Bussmann13}. 
    
\end{enumerate}

We present all of the CO measurements in Fig.\,\ref{fig:lcow50}. The various populations show a similar distribution in line width with a median around 300\,\kms, except the SMGs, which show a $\sim0.2$\,dex offset to higher velocities. On the other hand, the line luminosities increase with redshift, as expected from the limited instrument sensitivity. Even without correcting the line widths for the inclination angles, the correlation between \lco\ and $w$ is evident within each galaxy population. This is expected from the TFR if the galaxies in each population have similar molecular gas fraction ($f_{\rm mol} \equiv M_{\rm mol}/{M_{\rm bary}}$) and \cohh\ conversion factor ($\alpha_{\rm CO} \equiv M_{\rm mol}/L^{\prime}_{\rm CO}$). The dashed lines in Fig.\,\ref{fig:lcow50} show the expected correlations at a range of fixed $f_{\rm mol}/\aco$ ratios, and they seem to fit the data points well if we allow $f_{\rm mol}/\aco$ to vary among populations. This hints at an evolution that we will explore in detail in the next sections.  

\section{Analysis Method} \label{sec:analysis}

\begin{figure}
\includegraphics[width=0.48\textwidth]{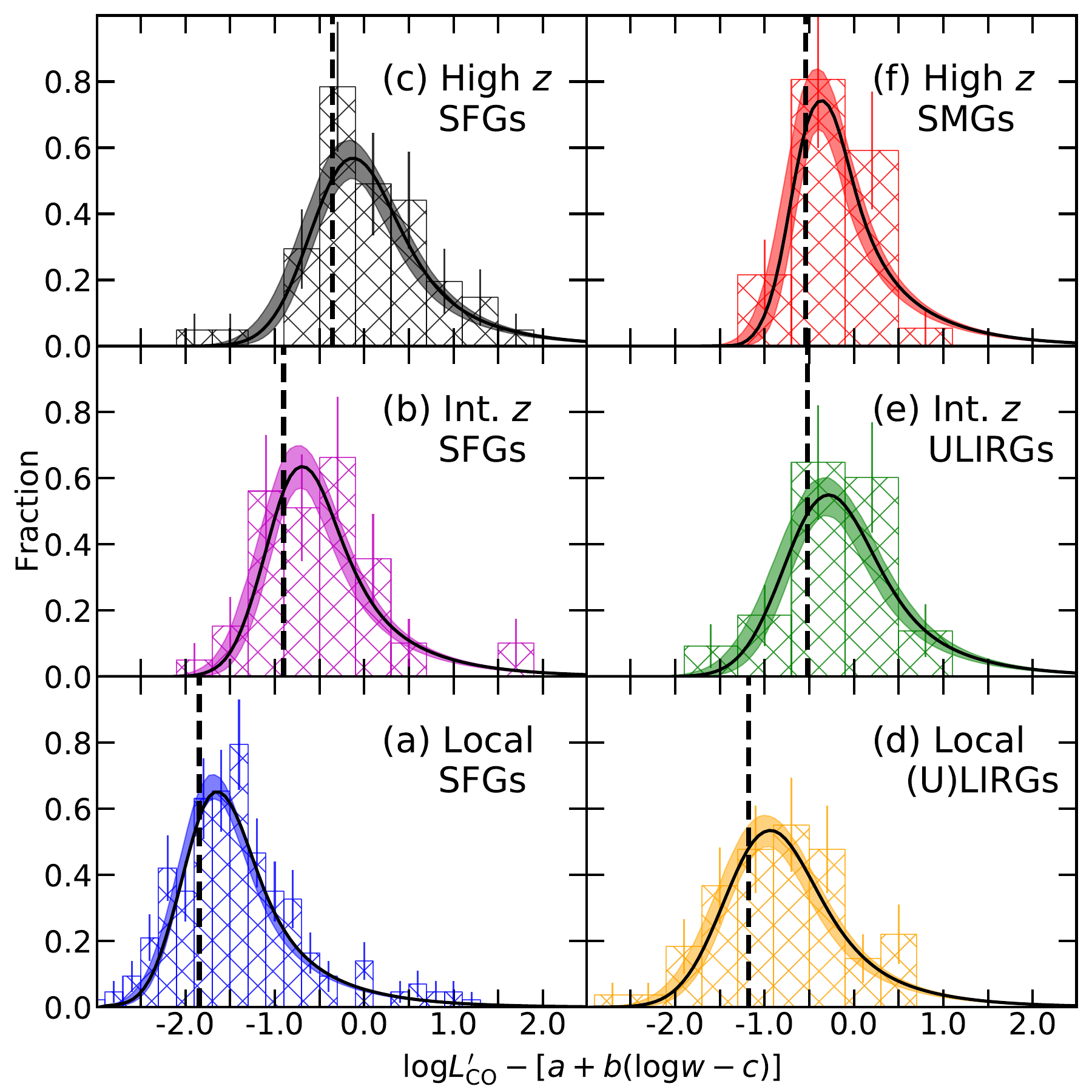}
\caption{CO data vs. best-fit model for normal star-forming populations ({\it left panels}) and starburst populations ({\it right panels}). The histograms show the distributions of the vertical offsets of the data points from the left-most dashed line in Fig.\,\ref{fig:lcow50}, which is calculated as $\log L^{\prime}_{\rm CO} - [a+b(\log w -c)]$ (i.e. the left side of Eq.\,\ref{eq:rsum}). The black solid curves show the probability density function (PDF) of $\log \fpm + X$ (i.e., the right side of Eq.\,\ref{eq:rsum}) using the best-fit parameters from the MCMC chains. The color-shaded regions around each curve show the $1\sigma$ confidence intervals of the PDFs, and the dashed vertical lines indicate the best-fit $\log \fpm$. To compare with the PDFs, the histograms are plotted in units of fraction per dex for 0.2-0.6\,dex bins. The error bars indicate Poisson noise. These histograms are shown for illustration only and are not used in the inference of best-fit parameters.}
\label{fig:pdffit}
\end{figure}

The CO line luminosity \lco\ traces the molecular gas mass $M_{\rm mol}$ through the \cohh\ conversion factor, while the inclination-corrected line width ($W \equiv w/\sin\,i$ where $i=0$ is face-on) provides a measure of the baryonic mass ${M_{\rm bary}}$ through the Tully-Fisher relation. Therefore, each CO measurement pair of \lco\ and $W$ offers an estimate of the ratio between the molecular gas fraction ($f_{\rm mol}$) and the \cohh\ conversion factor ($\aco$):
\begin{equation}
\frac{L^{\prime}_{\rm CO}}{{M_{\rm bary}}} = \frac{L^{\prime}_{\rm CO}}{M_{\rm mol}} \frac{M_{\rm mol}}{{M_{\rm bary}}} = \frac{f_{\rm mol}}{\alpha_{\rm CO}} \equiv \fpm
\label{eq:gco}
\end{equation}
For simplicity, we have defined the above ratio as $\fpm$, the $\aco$-normalized molecular gas fraction. Our goal is to measure the average $\fpm$ values of different galaxy populations as a function of redshift, from only a pair of CO-based observables. The detailed procedures are described below.

To estimate ${M_{\rm bary}}$, we adopt the CO baryonic TFR in the form of
\begin{equation}
\log\left(\frac{M_{\rm bary}}{M_{\odot}}\right) = a + b \left[\log\left(\frac{W}{{\rm km~s}^{-1} }\right)-c\right],
\label{eq:tfr}
\end{equation}
where $W$ is the inclination-corrected CO line width, i.e., $W=w/\sin\,i$.
For comparison, the CO {\it stellar-mass} TFR derived by \citet{Tiley:2016aa} is
\begin{multline}
\log\left(\frac{\mstar}{M_{\odot}}\right) = 10.51\pm0.04 \\
+ (3.3\pm0.3) \left[\log\left(\frac{W}{{\rm km~s}^{-1}}\right)-2.58\right],
\label{eq:stfr}
\end{multline}
which is calibrated using a large sample of local SFGs from COLD GASS \citep{Saintonge11,Saintonge17}. By incorporating the atomic gas mass from GASS \citep{Catinella:2018aa} and ALFALFA \citep[][]{Haynes18}, and the stellar mass and the molecular gas mass from COLD GASS, we can estimate the average stellar mass fraction, $f_{\star} \equiv \mstar/{M_{\rm bary}}=\mstar/(\mstar + M_{\rm HI} + M_{\rm mol})$, for the COLD GASS galaxies. As expected, these galaxies are dominated by stellar mass with $\langle f_{\star}\rangle\simeq 80\%$, which indicates that this {\it stellar-mass} TFR is a good approximation of the local {\it baryonic} TFR. We thus apply a small baryonic-mass correction ($1/f_\star = 1.25$ or 0.1\,dex) to Eq.\,\ref{eq:stfr} to obtain the coefficients of the CO baryonic TFR:
\begin{equation}
\begin{aligned}
a &= 10.61\pm0.04\\
b &= 3.3\pm0.3\\
c &= 2.58.
\end{aligned}
\end{equation}

The advantage of a baryonic TFR is that the same locally calibrated relation may still be valid at higher redshifts, because neither the dynamical equilibrium physics nor the baryonic-to-dark-matter mass ratio of halos is expected to strongly evolve with redshift \citep{McGaugh:2012aa}. In contrast, the stellar-mass TFR evolves with redshift \citep{Cresci:2009aa,Miller:2011aa,Miller:2012aa}, likely because of the increase in gas fraction with redshift. 

To utilize the TFR for the mass estimation of individual galaxies, one must correct the line width for the inclination angle of the disk, which is not always available especially at high redshifts due to limited spatial resolution. Therefore, we correct the inclination angles statistically, by considering the galaxies in each population as randomly oriented disks. We can determine the average $\fpm$ of the population by matching the distribution of observables with the expected probability density function (PDF). 

We begin by expressing the relations between the observed properties $(w, L^{\prime}_{\rm CO})$ and their true values as

\begin{equation}
\begin{aligned}
\log w &= \log(w_{\rm true}) + \epsilon_w \\ \log L^{\prime}_{{\rm CO}} &=  \log L^{\prime}_{{\rm CO, true}} + \epsilon_{L}.
\end{aligned}
\label{eq:wL}
\end{equation}
Here, $w_{\rm true} \equiv W_{\rm true} \sin i$ is the true line FWHM. The random variables ($\epsilon_w$ and $\epsilon_L$) represent the fractional measurement errors:
\begin{equation}
\begin{aligned}
\epsilon_{w} & \equiv \log \frac{w}{w_{{\rm true}}} \simeq \frac{1}{\ln 10}\frac{w-w_{{\rm true}}}{w_{{\rm true}}}  \\
\epsilon_{L} & \equiv \log \frac{L^{\prime}_{{\rm CO}}}{L^{\prime}_{{\rm CO, true}}} \simeq \frac{1}{\ln 10} \frac{L^{\prime}_{{\rm CO}}-L^{\prime}_{{\rm CO, true}}}{L^{\prime}_{{\rm CO, true}}}.
\end{aligned}
\label{eq:epsilon}
\end{equation}
Next, we rewrite the $\aco$-normalized molecular gas fraction $\fpm$ defined in Eq.\,\ref{eq:gco} using the relations in Eq.\,\ref{eq:wL} and the TFR in Eq.\,\ref{eq:tfr} as
\begin{equation}
\begin{aligned}
\log \fpm & = \log L^{\prime}_{{\rm CO, true}} - \log M_{{\rm bary, true}} \\
	& = (\log {L'_{{\rm CO}}} - \epsilon_{L}) - [a + b (\log{w}-\log(\sin\,i)-\epsilon_{w}-c)].
\end{aligned}
\end{equation}
Finally, we can rearrange the above equation and get
\begin{equation}
\begin{aligned}
  \log {L'_{{\rm CO}}} - [a + b(\log w-c)] 
   &= \log \fpm + \epsilon_{L} - b\epsilon_{w} - b\log(\sin\,i) \\
   &\equiv \log \fpm + X,
\label{eq:rsum}
\end{aligned}
\end{equation}
in which the left side is a combined observable that can be determined from the CO measurement pair $(w, L^{\prime}_{{\rm CO}})$, while the right side is the sum between $\log \fpm$ and a linear combination of three random variables, expressed as $X\equiv\epsilon_L-b\epsilon_w-b\log(\sin i)$.
If the PDFs of these random variables are known, one can estimate $\log \fpm$ by matching the expected PDF of $\log \fpm + X$ to the observed distribution of $\log {L'_{{\rm CO}}} - [a + b(\log w-c)]$ for a galaxy population. 

As a linear combination of three independent random variables, the PDF of $X$ is the convolution of their individual PDFs:
\begin{eqnarray}
f_X(x)= f_{\epsilon_L}(x) * f_{-b\epsilon_{w}}(x) * f_{-b\log(\sin\,i)}(x).
\end{eqnarray}
Because $\epsilon_{w}$ and $\epsilon_{L}$ represent fractional measurement errors (Eq.\,\ref{eq:epsilon}), we can assume that they are drawn from two Gaussian distributions with dispersions of $\sigma_w$ and $\sigma_L$, respectively. Their convolution is still a Gaussian, and we have the PDF of ($\epsilon_L - b\epsilon_{w}$):
\begin{eqnarray}\label{eq:mserr}
f_{ \epsilon_L - b\epsilon_{w} } (x)  & = &  \frac{1}{{\sigma \sqrt {2\pi } }}e^{{{ -  {x}^2 } / {2\sigma^2 }}}\\
\sigma^2 & = & b^2\sigma_w^2+\sigma_L^2.
\end{eqnarray}
On the other hand, given the PDF of the inclination angle $i$ for random orientations, $f_i(x) = \sin x$ for $0 \leq x \leq \pi/2$, we derive the PDF of $-b\log(\sin\,i)$:
\begin{equation}\label{eq:sini}
        f_{-b\log(\sin\,i)}(x)=
        \begin{dcases}
            \frac{\ln 10}{b} \frac{10^{-2x/b}}{\sqrt{1-10^{-2x/b}}} & \quad x \geq 0 \\
            0 & \quad x < 0
        \end{dcases}
\end{equation}
Its convolution with Eq.\,\ref{eq:mserr} gives the expected PDF of $X$:
\begin{equation}
f_X(x) = \frac{\ln 10}{b} \int_{0}^{\infty}\frac{10^{-2t/b}}{\sqrt{1-10^{-2t/b}}}
\frac{1}{{\sigma \sqrt {2\pi } }}e^{{ -  (x-t)^2 } / {2\sigma ^2 }} dt,
\label{eq:fx}
\end{equation}
which peaks near zero. As a result, the PDF of $\log \fpm + X$ peaks near $\log \fpm$. 

To obtain the model parameters ($\fpm$, $\sigma$) that best describe the data, we write down the likelihood function of the observed data set $\{w_k, L'_{{\rm CO},k}\}$ given a model described by $\fpm$ and $\sigma$ as
\begin{equation}
\mL(\fpm,\sigma) \equiv p(\{w_k, L'_{{\rm CO},k}\} | \fpm, \sigma) = \prod_{k} f_X(x_k),
\end{equation}
where $f_X$ is from Eq.\,\ref{eq:fx} and $x_k$ is calculated for the $k$-th galaxy in the population:
\begin{eqnarray}
x_k=\log {L'_{{\rm CO},k}} - [a + b(\log w_k-c)] - \log \fpm.
\end{eqnarray}
This likelihood function is maximized when the model parameters best describe the observed data set. Using Bayes' Theorem, the posterior PDF of the model given the data, $p(\fpm, \sigma | \{w_k, L'_{{\rm CO},k}\})$, is the product of the likelihood function and the model prior $p(\fpm, \sigma)$:
\begin{equation}
p(\fpm, \sigma |\{w_k, L'_{{\rm CO},k}\}) \propto \mL (\fpm, \sigma) p( \fpm, \sigma).
\label{eq:postpdf}
\end{equation}
To sample the posterior PDFs and quantify the best-fit values of ($\fpm$, $\sigma$) and their uncertainties, we use the Affine Invariant Markov chain Monte Carlo (MCMC) Ensemble sampler implemented in Python code \emcee\footnote{\url{http://dfm.io/emcee}} \citep{Foreman-Mackey:2013aa}. We assume bounded ``flat'' priors for both $\log \fpm$ and $\sigma$: $-10 \leq \log \fpm \leq 10$ and $0 \leq \sigma \leq 10$. 

\section{Results} \label{sec:result}

In Table\,\ref{tab:cats}, we report the the median values of the marginalized posterior distributions from the MCMC chains as the best-fit parameters and the 15.8-and-84.1-percentiles as the 1$\sigma$ confidence intervals. To illustrate how well our models describe the data, in Fig.\,\ref{fig:pdffit} we compare the model PDFs using the best-fit parameters from \emcee\ and the observed distribution of $\log {L'_{{\rm CO}}} - [a + b(\log w-c)]$ for each galaxy population. For all six populations, the model PDFs fit the histograms quite well, validating the \emcee\ results.     

\subsection{The Evolution of Molecular Gas Fraction}

\begin{figure}
\includegraphics[width=0.45\textwidth]{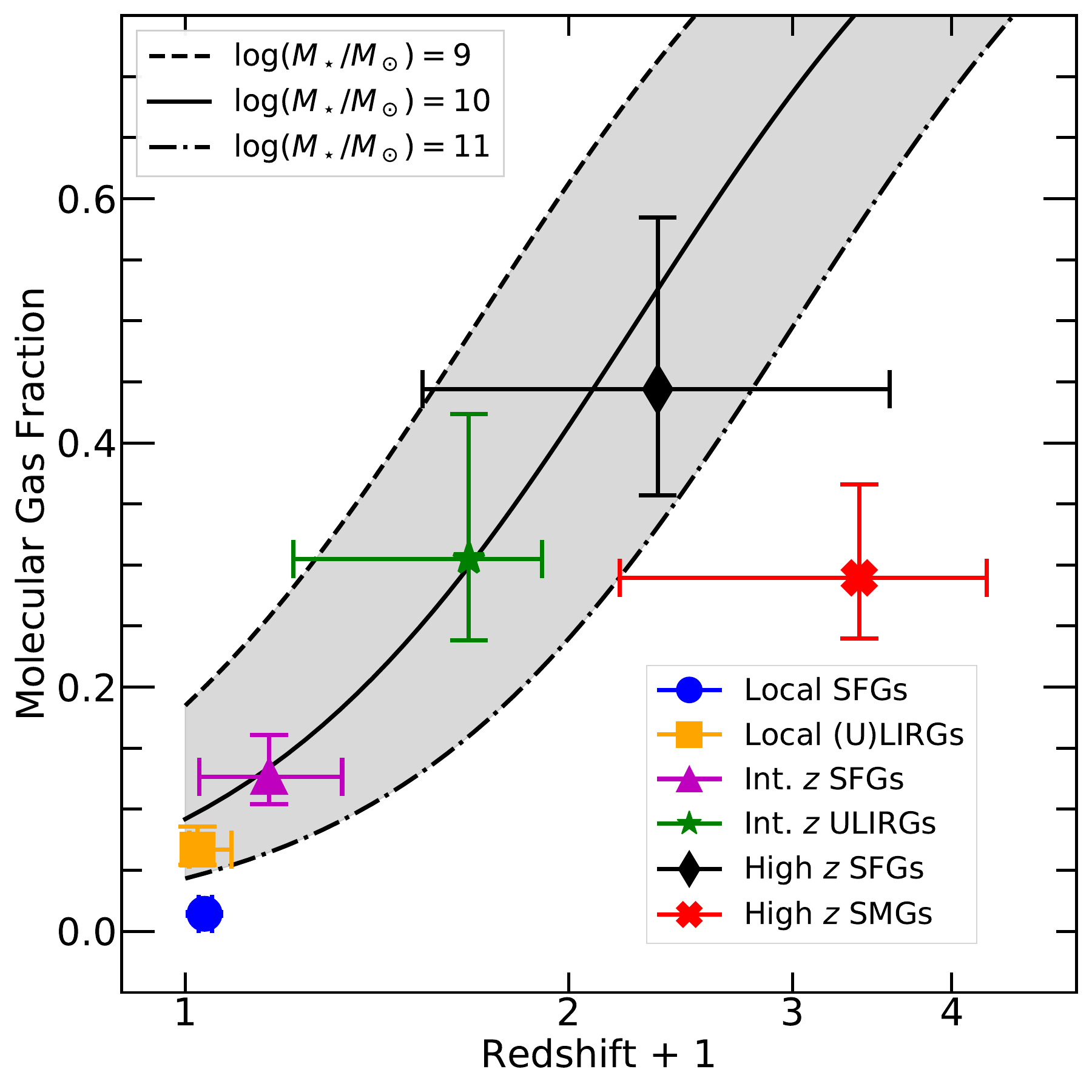}
\caption{The redshift evolution of the molecular gas fraction. The plotted $f_{\rm mol}$ values and their uncertainties are calculated from $\log \fpm$ assuming $\alpha_{\rm CO} = 1.0\,M_\odot/L_l$ for all populations. Data points are plotted at the median redshift of each population with horizontal error bars representing the redshift range. For comparison, the dashed, solid, and dash-dotted curves show the inferred $f_{\rm mol}$ evolution for stellar masses of $10^9, 10^{10}$, and $10^{11}$\,\msun, respectively, based on the observed evolution of the star-forming main sequence and a  Kennicutt-Schmidt relation (Eqs.\,\ref{eq:kslaw} and \ref{eq:ssfr}).} 
\label{fig:fmolz}
\vskip 0.1in
\end{figure}

In the previous section, we have shown that the CO measurements alone can provide an estimate of the mean molecular gas fraction for a galaxy population, given a baryonic TFR and a \cohh\ conversion factor. The histograms and best-fit models in Fig.\,\ref{fig:pdffit} clearly show a redshift evolution of the $\aco$-normalized molecular gas fraction ($\fpm$), as highlighted by the offsets in their peaks. To better illustrate this redshift evolution, we plot $f_{\rm mol}$ as a function of redshift in Fig.\,\ref{fig:fmolz} for the six galaxy populations compiled in Table\,\ref{tab:cats}. To convert $\fpm$ to $f_{\rm mol}$, we have applied a fiducial $\alpha_{\rm CO}$ value of $1\,M_\odot/L_l$ for all populations in both Fig.\,\ref{fig:fmolz} and in Table\,\ref{tab:cats}. The redshift evolution of the gas fraction is evident, and it roughly follows a power-law $f_{\rm mol} \propto (1+z)^\beta$ with $\beta \sim 2$ between $0 < z < 3$. 

As a consistency check, we compare our results with the observed redshift evolution of the star-forming main sequence. The molecular gas fraction can be inferred from the normalization of the star-forming main sequence (i.e., the specific SFR, sSFR = SFR/$\mstar$) and the Kennicutt-Schmidt star formation relation (SFR = $M_{\rm mol}/\tau$, where $\tau$ is the gas depletion timescale) because
\begin{equation}
f_{\rm mol} \simeq \frac{M_{\rm mol}}{M_{\rm mol} + \mstar} = \frac{{\rm sSFR}\cdot\tau}{1+{\rm sSFR}\cdot\tau}.
\label{eq:kslaw}
\end{equation}
For main-sequence SFGs, it is appropriate to use a gas depletion timescale of $\tau \sim 0.7$\,Gyr \citep[e.g.,][]{Tacconi13,Saintonge17}\footnote{Note that \citet{Tacconi13} inferred a $\sim$2$\times$ longer gas depletion timescale ($\tau = 1.5$\,Gyr at $z \sim 0$) from COLD GASS because the SFR at $M_\star \sim 10^{10.5}$\,\msun\ used in \citet{Saintonge17} is $\sim$2$\times$ lower than the best-fit SFR from Eq.\,\ref{eq:ssfr}. For consistency, we adopt $\tau = 0.7$\,Gyr at all redshifts.}. The observed sSFR of main-sequence SFGs depends strongly on redshift and mildly on stellar mass, and the best-fit polynomial function is \citep[see references in][for the original data]{Tacconi13}:
\begin{equation}
{\rm sSFR} = 0.68\,{\rm Gyr}^{-1} \Big(\frac{\mstar}{6.6\times10^{10}\,M_\odot}\Big)^{-0.35} \Big(\frac{1+z}{2.2}\Big)^{2.8}.
\label{eq:ssfr}
\end{equation}
Using the above two relations, we can infer $f_{\rm mol}$ at any given redshift and stellar mass. The curves and shaded areas in Fig.\,\ref{fig:fmolz} show the inferred evolution of $f_{\rm mol}$ at fixed stellar masses of $\mstar = 10^9, 10^{10}, 10^{11}$\,\msun. 
Without making any adjustments, our results closely follow the trend inferred from the observed evolution of the main sequence in Fig.\,\ref{fig:fmolz}. Almost all of the data points follow the curve for a stellar masses between $10^{10} < \mstar < 10^{11}$\,\msun, except for the local SFGs and the high-$z$ SMGs. 

For the local SFGs, the plotted $f_{\rm mol}$ lies significantly below the shaded area because a higher Galactic-like \cohh\ conversion factor is more appropriate for these galaxies. Using the Milky-Way value of $\alpha_{\rm CO} = 4.3\,M_\odot/L_l$, $f_{\rm mol}$ increases from $1.4\%$ to $6.0\%$, approaching the shaded area in Fig.\,\ref{fig:fmolz} and becoming consistent with that of local (U)LIRGs ($6.7^{+1.3}_{-1.7}\%$). By design, this is in perfect agreement with the mean molecular gas fraction of $f_{\rm mol} \simeq 6\%$ for the same COLD GASS sample, based on a direct calculation using their stellar masses, \ion{H}{1} masses, and molecular masses ($M_{\rm mol}/(\mstar + M_{\rm HI} + M_{\rm mol})$). But on the other hand, if we assume that local SFGs and local (U)LIRGs should have similar molecular fractions, then the best-fit $\fpm$ values would {\it indicate} that the former should have a $4.8\pm1.2$ times greater $\aco$ value than the latter. This is consistent with the findings from previous $\alpha_{\rm CO}$ studies of local (U)LIRGs \citep[e.g.,][]{Downes:1998aa,Papadopoulos:2012ab}.

Our model shows that the SMGs have a mean molecular gas fraction of $f_{\rm mol} = 28^{+5}_{-7}$\% for $\aco = 1.0\,M_\odot/L_l$ at $\overline{z} \sim 2.5$. Similar to local SFGs, their data point lies significantly below the solid curve in Fig\,\ref{fig:fmolz}. But for the SMGs, the adopted starburst-like \cohh\ conversion factor is supported by other observations \citep[e.g.,][]{Magdis11,Hodge12,Magnelli12b,Xue18}. There are two possible ways to explain this result. First, the SMGs contain a large fraction of mergers that are spatially unresolved in the CO observations \citep[e.g.,][]{Engel10,Fu12b,Fu13}. The relative velocities between merging components increase the line widths and thus decrease the molecular gas fraction inferred from the TFR. Second, the SMGs may have stellar masses exceeding $10^{11}$\,\msun, which is higher than the stellar mass of high-$z$ SFGs from the PHIBSS survey \citep[$\mstar \sim 10^{10.5}$\,\msun;][]{Tacconi13}. The mean stellar mass of SMGs is still a matter of debate, with estimates between $\sim5\times10^{10}$\,\msun\ \citep{Hainline11} and $\sim3\times10^{11}$\,\msun\ \citep[e.g.,][]{Michaowski10,Michalowski12}. Our current estimate is more consistent with the higher estimate. 

Lastly, if we were to adopt the higher Galactic-like \cohh\ conversion factor on the high-$z$ SFGs, $f_{\rm mol}$ would exceed 100\%, which is unphysical. We thus obtain an upper limit on the \cohh\ conversion factor of $\alpha_{\rm CO,SFG} < 2.3^{+0.9}_{-0.4}\,M_\odot/L_l$ for SFGs at $\overline{z} \sim 1.5$. Similarly, for SMGs, the upper limit is at $\alpha_{\rm CO,SB} < 3.5^{+1.1}_{-0.6}\,M_\odot/L_l$.

\subsection{Systematic Uncertainties from the TFR} \label{sec:sysunc}

In the above analysis, we have ignored the uncertainties of the TFR itself, which include uncertainties of the zero-point mass (parameter $a$), the slope (parameter $b$), and the intrinsic scattering ($\sigma_{\rm int}$). 

\citet{Tiley:2016aa} quoted 1$\sigma$ uncertainties of 0.04\,dex in $a$ and 0.3 in $b$ (which is also 0.04\,dex because $b = 3.3$). We examine the systematic uncertainties of the best-fit $\fpm$ values by varying the coefficients of the TFR and repeating the analysis. We find that the above uncertainties of the $a$ and $b$ coefficients translate to systematic uncertainties of $\fpm$ around 0.04, 0.06, and 0.1\,dex for the local, intermediate-$z$, and high-$z$ samples, respectively. In all cases, the systematic uncertainty is comparable to {\it or} smaller than the statistical errors from the MCMC chains. 

The measured TFR zero-point mass and the slope are also known to vary among different studies, depending on their choices of the kinematics and the mass tracers. For example, the TFRs based on the asymptotic rotation velocity or the velocity along the flat part of a resolved rotation curve typically exhibit steeper slope ($b \sim 4$) \citep[e.g.,][]{Miller:2011aa,McGaugh:2015aa,Papastergis:2016aa} than those based on line width \citep[e.g.,][]{McGaugh:2012aa,Tiley:2016aa}. Additionally, the TFRs using \ion{H}{1}, H$\alpha$, [\ion{O}{2}] may differ from those using CO, because the kinematics tracers could sample different spatial scales in a galaxy. 
We chose the CO TFR from \citet{Tiley:2016aa} to minimize the impact from these systematics because it uses the width of a low-$J$ CO line, similar to the application in this study.
Nevertheless, because varying the TFR coefficients would change the estimated $\fpm$ values of all galaxy populations along the same direction, the observed redshift evolution of $\fpm$ is unchanged. 

At a given line width, an intrinsic scatter of $\sigma_{\rm int}=0.1-0.2$\,dex in mass is expected in the baryonic TFR, due to the variations in the mass concentration relation of dark matter halos and the baryonic-to-halo mass ratio \citep[e.g.,][]{Dutton:2012aa}. Because the intrinsic scatter of the TFR affects the observables in the same way as measurement errors, the dispersion parameter $\sigma$ of our model should include the contribution from $\sigma_{\rm int}$; specifically, $\sigma^2 = b^2 \sigma_w^2 + \sigma_L^2 + \sigma_{\rm int}^2$. But the observational uncertainties are not qualified accurately enough to separate the measurement errors and the intrinsic scatter in $\sigma$. The expected intrinsic scatter of $0.1-0.2$\,dex is significantly smaller than what we measured from the data ($\sigma \simeq 0.4-0.6$\,dex; see Table\,\ref{tab:cats}), which appear to be dominated by the measurement error of line width ($b \sigma_w \sim 0.35$\,dex for a typical error of 25\%). 
 
\section{Summary and Future Prospects} \label{sec:summary}

In summary, we have developed a new method to infer the mean molecular gas fraction of a galaxy population. This method requires only spatially-integrated low-$J$ CO observations and corrects the inclination effects statistically. This is possible because (1) the CO line luminosity traces with the molecular gas mass through a \cohh\ conversion factor; (2) the CO line width, once corrected for the disk inclination angle, provides the total baryonic mass, ${M_{\rm bary}}$, given a baryonic CO TFR; and (3) the ratio of the two, $L^{\prime}_{\rm CO} / {M_{\rm bary}}$, is the $\aco$-normalized molecular gas fraction, defined as $\fpm \equiv f_{\rm mol}/\alpha_{\rm CO}$. 
We use the expected PDF of the inclination angle from randomly oriented disks to correct the inclination effect statistically. The model also accounts for the measurement errors and the intrinsic dispersion of the TFR in a dispersion parameter $\sigma$. From the literature, we have compiled CO measurements for three populations of normal SFGs and three populations of starburst galaxies with redshifts stretching between $0.01 < z < 3.26$. We use Bayesian inference and the MCMC sampler \emcee\ to derive the joint and marginalized PDFs for $\fpm$ and $\sigma$ from the CO data of each populations. Our main findings are as follows:

\begin{enumerate}

\item The molecular gas fraction increases rapidly with redshift for both normal SFGs and starbursts. The evolution trend is consistent with that indirectly inferred from the observed evolution of the star-forming main sequence and a Kennicutt-Schmidt relation; 

\item By comparing the inferred $\fpm$ values for the local SFGs and local (U)LIRGs, we find that the two populations would have similar molecular gas fractions only if a $\sim$5$\times$ higher $\alpha_{\rm CO}$ conversion factor is used for the former population, consistent with previous $\alpha_{\rm CO}$ studies of local galaxies;

\item At higher redshifts ($z > 1$), our results suggest a lower starburst-like $\aco$ is applicable for both main-sequence galaxies and starbursts. In fact, the upper limit of molecular gas fraction at 100\% translates to upper limits of $\aco$ for these populations: $\alpha_{\rm CO,SFG} <2.3^{+0.9}_{-0.4}\,M_\odot/L_l$ and $\alpha_{\rm CO, SB}<3.5^{+1.1}_{-0.6}\,M_\odot/L_l$ for high-$z$ SFGs and starbursts, respectively;

\item The molecular gas fraction of SMGs is relatively low compared to that of coeval main-sequence galaxies, indicating a significant fraction of unresolved mergers and/or an average stellar mass exceeding $10^{11}$\,\msun.

\end{enumerate}

Clearly, our results hinge upon the assumption that the baryonic CO TFR does not evolve with redshift. Galaxy formation models suggest weak evolution of the baryonic TFR utilizing the maximum circular velocity ($v_{\rm max}$), because individual galaxies evolve along such scaling relations \citep[e.g.,][]{Dutton:2011aa}. To extend the theoretical expectation to the CO TFR we adopted, we have implicitly assumed a nearly constant ratio between $v_{\rm max}$ and the CO FWHM. This assumption can be tested with future spatially resolved measurements of rotation curves in a large sample of galaxies across a wide redshift range. 

While our choice of the CO lines was motivated by their availability in the literature, they are also favored for several other reasons. Firstly, they allow us to calibrate the TFR with local galaxies and apply it at higher redshifts with the same kinematic tracer (CO FWHM), easing the concern of the systematic biases introduced by different kinematic tracers \citep[see][]{Bradford:2016aa}. Secondly, we have restricted the sample to galaxies with low-$J$ CO measurements, which are expected to have a larger spatial extent than high-$J$ CO emission. In both local and high-$z$ galaxies, we expect that a substantial fraction of the low-$J$ CO emission reaches the flat part of the rotation curve \citep[e.g.,][]{Downes:1998aa,Hodge12,Xue18,Pereira-Santaella2018:aa}. Lastly, given the higher gas fraction and larger molecular-to-atomic ratio at high redshifts \citep[e.g.,][]{Lagos:2011ab}, we would expect that the low-$J$ CO lines are even more effective at tracing global kinematics at high redshifts than in nearby galaxies.

The likelihood method presented here can also be used to measure TFRs without knowing the inclination angles of individual galaxies, which is particularly useful for high-$z$ galaxies. Similar to the ``inclination-free'' maximum likelihood estimation method of \citet{Obreschkow:2013aa}, a galaxy sample only needs to have measurements of the mass (either stellar or baryonic mass) and the line width to measure the TFR. This Bayesian-based analysis can yield reliable measurements of the TFR if the sample contains enough objects (ideally $N > 50$) and has roughly uniform measurement errors. 

Future studies of the molecular gas fraction will certainly benefit from more and better CO data from large surveys. The Atacama Spectroscopic Survey \citep[ASPECS;][]{Walter16} is a good start toward a wide-field blind CO survey, while the next generation Very Large Array (ngVLA) will further probe \textit{unlensed} CO\,($1\rightarrow0$) transitions at $z>1.5$ \citep{Emonts:2018aa,Decarli:2018aa}. In addition, one can apply the same method on the original TFR tracer, the \hi\ 21\,cm line, to study the atomic gas fraction evolution. Such a study will be viable when \hi\ detections of large samples of galaxies become available up to $z \simeq 0.6$ with the VLA \citep[e.g., ][]{Fernandez:2016aa} and up to $z \simeq 1.4$ with the Square Kilometer Array \citep[e.g.,][]{Booth:2009aa,Allison:2015aa}.

\acknowledgements 

We thank the anonymous referee for helpful comments and acknowledge support from the National Science Foundation (NSF) grant AST-1614326.

\bibliographystyle{aasjournal}

\begin{thebibliography}{}
\expandafter\ifx\csname natexlab\endcsname\relax\def\natexlab#1{#1}\fi
\providecommand{\url}[1]{\href{#1}{#1}}
\providecommand{\dodoi}[1]{doi:~\href{http://doi.org/#1}{\nolinkurl{#1}}}
\providecommand{\doeprint}[1]{\href{http://ascl.net/#1}{\nolinkurl{http://ascl.net/#1}}}
\providecommand{\doarXiv}[1]{\href{https://arxiv.org/abs/#1}{\nolinkurl{https://arxiv.org/abs/#1}}}

\bibitem[{{Allison} {et~al.}(2015){Allison}, {Sadler}, {Moss}, {Whiting},
  {Hunstead}, {Pracy}, {Curran}, {Croom}, {Glowacki}, {Morganti}, {Shabala},
  {Zwaan}, {Allen}, {Amy}, {Axtens}, {Ball}, {Bannister}, {Barker}, {Bell},
  {Bock}, {Bolton}, {Bowen}, {Boyle}, {Braun}, {Broadhurst}, {Brodrick},
  {Brothers}, {Brown}, {Bunton}, {Cantrall}, {Chapman}, {Cheng}, {Chippendale},
  {Chung}, {Cooray}, {Cornwell}, {DeBoer}, {Diamond}, {Edwards}, {Ekers},
  {Feain}, {Ferris}, {Forsyth}, {Gough}, {Grancea}, {Gupta}, {Guzman},
  {Hampson}, {Harvey-Smith}, {Haskins}, {Hay}, {Hayman}, {Heywood}, {Hotan},
  {Hoyle}, {Humphreys}, {Indermuehle}, {Jacka}, {Jackson}, {Jackson},
  {Jeganathan}, {Johnston}, {Joseph}, {Kendall}, {Kesteven}, {Kiraly},
  {Koribalski}, {Leach}, {Lenc}, {Lensson}, {Mackay}, {Macleod}, {Marquarding},
  {Marvil}, {McClure-Griffiths}, {McConnell}, {Mirtschin}, {Norris}, {Neuhold},
  {Ng}, {O'Sullivan}, {Pathikulangara}, {Pearce}, {Phillips}, {Popping},
  {Qiao}, {Reynolds}, {Roberts}, {Sault}, {Schinckel}, {Serra}, {Shaw},
  {Shields}, {Shimwell}, {Storey}, {Sweetnam}, {Troup}, {Turner}, {Tuthill},
  {Tzioumis}, {Voronkov}, {Westmeier}, \& {Wilson}}]{Allison:2015aa}
{Allison}, J.~R., {Sadler}, E.~M., {Moss}, V.~A., {et~al.} 2015, \mnras, 453,
  1249, \dodoi{10.1093/mnras/stv1532}

\bibitem[{Armus {et~al.}(2009)Armus, Mazzarella, Evans, Surace, Sanders,
  Iwasawa, Frayer, Howell, Chan, Petric, Vavilkin, Kim, Haan, Inami, Murphy,
  Appleton, Barnes, Bothun, Bridge, Charmandaris, Jensen, Kewley, Lord, Madore,
  Marshall, Melbourne, Rich, Satyapal, Schulz, Spoon, Sturm, U, Veilleux, \&
  Xu}]{Armus09}
Armus, L., Mazzarella, J.~M., Evans, A.~S., {et~al.} 2009, \pasp, 121, 559

\bibitem[{{Bigiel} {et~al.}(2008){Bigiel}, {Leroy}, {Walter}, {Brinks}, {de
  Blok}, {Madore}, \& {Thornley}}]{Bigiel:2008aa}
{Bigiel}, F., {Leroy}, A., {Walter}, F., {et~al.} 2008, \aj, 136, 2846,
  \dodoi{10.1088/0004-6256/136/6/2846}

\bibitem[{{Blitz}(1993)}]{Blitz:1993aa}
{Blitz}, L. 1993, in Protostars and Planets III, ed. E.~H. {Levy} \& J.~I.
  {Lunine}, 125--161

\bibitem[{{Bolatto} {et~al.}(2013){Bolatto}, {Wolfire}, \& {Leroy}}]{Bolatto13}
{Bolatto}, A.~D., {Wolfire}, M., \& {Leroy}, A.~K. 2013, \araa, 51, 207,
  \dodoi{10.1146/annurev-astro-082812-140944}

\bibitem[{{Booth} {et~al.}(2009){Booth}, {de Blok}, {Jonas}, \&
  {Fanaroff}}]{Booth:2009aa}
{Booth}, R.~S., {de Blok}, W.~J.~G., {Jonas}, J.~L., \& {Fanaroff}, B. 2009,
  ArXiv e-prints.
\newblock \doarXiv{0910.2935}

\bibitem[{{Bothwell} {et~al.}(2013){Bothwell}, {Maiolino}, {Kennicutt},
  {Cresci}, {Mannucci}, {Marconi}, \& {Cicone}}]{Bothwell13a}
{Bothwell}, M.~S., {Maiolino}, R., {Kennicutt}, R., {et~al.} 2013, \mnras, 433,
  1425, \dodoi{10.1093/mnras/stt817}

\bibitem[{Bothwell {et~al.}(2013)Bothwell, Smail, Chapman, Genzel, Ivison,
  Tacconi, Alaghband-Zadeh, Bertoldi, Blain, Casey, Cox, Greve, Lutz, Neri,
  Omont, \& Swinbank}]{Bothwell13}
Bothwell, M.~S., Smail, I., Chapman, S.~C., {et~al.} 2013, \mnras, 429, 3047

\bibitem[{Bouch{\'e} {et~al.}(2010)Bouch{\'e}, Dekel, Genzel, Genel, Cresci,
  F{\"o}rster~Schreiber, Shapiro, Davies, \& Tacconi}]{Bouche10}
Bouch{\'e}, N., Dekel, A., Genzel, R., {et~al.} 2010, \apj, 718, 1001

\bibitem[{{Bradford} {et~al.}(2016){Bradford}, {Geha}, \& {van den
  Bosch}}]{Bradford:2016aa}
{Bradford}, J.~D., {Geha}, M.~C., \& {van den Bosch}, F.~C. 2016, \apj, 832,
  11, \dodoi{10.3847/0004-637X/832/1/11}

\bibitem[{{Bussmann} {et~al.}(2013){Bussmann}, {P{\'e}rez-Fournon}, {Amber},
  {Calanog}, {Gurwell}, {Dannerbauer}, {De Bernardis}, {Fu}, {Harris}, {Krips},
  {Lapi}, {Maiolino}, {Omont}, {Riechers}, {Wardlow}, {Baker}, {Birkinshaw},
  {Bock}, {Bourne}, {Clements}, {Cooray}, {De Zotti}, {Dunne}, {Dye}, {Eales},
  {Farrah}, {Gavazzi}, {Gonz{\'a}lez Nuevo}, {Hopwood}, {Ibar}, {Ivison},
  {Laporte}, {Maddox}, {Mart{\'{\i}}nez-Navajas}, {Michalowski}, {Negrello},
  {Oliver}, {Roseboom}, {Scott}, {Serjeant}, {Smith}, {Smith}, {Streblyanska},
  {Valiante}, {van der Werf}, {Verma}, {Vieira}, {Wang}, \&
  {Wilner}}]{Bussmann13}
{Bussmann}, R.~S., {P{\'e}rez-Fournon}, I., {Amber}, S., {et~al.} 2013, \apj,
  779, 25, \dodoi{10.1088/0004-637X/779/1/25}

\bibitem[{{Carilli} \& {Walter}(2013)}]{Carilli:2013aa}
{Carilli}, C.~L., \& {Walter}, F. 2013, \araa, 51, 105,
  \dodoi{10.1146/annurev-astro-082812-140953}

\bibitem[{{Catinella} {et~al.}(2018){Catinella}, {Saintonge}, {Janowiecki},
  {Cortese}, {Dav{\'e}}, {Lemonias}, {Cooper}, {Schiminovich}, {Hummels},
  {Fabello}, {Ger{\'e}b}, {Kilborn}, \& {Wang}}]{Catinella:2018aa}
{Catinella}, B., {Saintonge}, A., {Janowiecki}, S., {et~al.} 2018, \mnras, 476,
  875, \dodoi{10.1093/mnras/sty089}

\bibitem[{Chiu {et~al.}(2007)Chiu, Bamford, \& Bunker}]{Chiu07}
Chiu, K., Bamford, S.~P., \& Bunker, A. 2007, \mnras, 377, 806

\bibitem[{{Combes} {et~al.}(2011){Combes}, {Garc{\'{\i}}a-Burillo}, {Braine},
  {Schinnerer}, {Walter}, \& {Colina}}]{Combes11}
{Combes}, F., {Garc{\'{\i}}a-Burillo}, S., {Braine}, J., {et~al.} 2011, \aap,
  528, A124, \dodoi{10.1051/0004-6361/201015739}

\bibitem[{{Combes} {et~al.}(2013){Combes}, {Garc{\'{\i}}a-Burillo}, {Braine},
  {Schinnerer}, {Walter}, \& {Colina}}]{Combes13}
---. 2013, \aap, 550, A41, \dodoi{10.1051/0004-6361/201220392}

\bibitem[{{Cresci} {et~al.}(2009){Cresci}, {Hicks}, {Genzel}, {Schreiber},
  {Davies}, {Bouch{\'e}}, {Buschkamp}, {Genel}, {Shapiro}, {Tacconi},
  {Sommer-Larsen}, {Burkert}, {Eisenhauer}, {Gerhard}, {Lutz}, {Naab},
  {Sternberg}, {Cimatti}, {Daddi}, {Erb}, {Kurk}, {Lilly}, {Renzini},
  {Shapley}, {Steidel}, \& {Caputi}}]{Cresci:2009aa}
{Cresci}, G., {Hicks}, E.~K.~S., {Genzel}, R., {et~al.} 2009, \apj, 697, 115,
  \dodoi{10.1088/0004-637X/697/1/115}

\bibitem[{{Davis} {et~al.}(2011){Davis}, {Bureau}, {Young}, {Alatalo}, {Blitz},
  {Cappellari}, {Scott}, {Bois}, {Bournaud}, {Davies}, {de Zeeuw}, {Emsellem},
  {Khochfar}, {Krajnovi{\'c}}, {Kuntschner}, {Lablanche}, {McDermid},
  {Morganti}, {Naab}, {Oosterloo}, {Sarzi}, {Serra}, \& {Weijmans}}]{Davis11}
{Davis}, T.~A., {Bureau}, M., {Young}, L.~M., {et~al.} 2011, \mnras, 414, 968,
  \dodoi{10.1111/j.1365-2966.2011.18284.x}

\bibitem[{{Decarli} {et~al.}(2018){Decarli}, {Carilli}, {Casey}, {Emonts},
  {Hodge}, {Kohno}, {Narayanan}, {Riechers}, {Sargent}, \&
  {Walter}}]{Decarli:2018aa}
{Decarli}, R., {Carilli}, C., {Casey}, C., {et~al.} 2018, ArXiv e-prints.
\newblock \doarXiv{1810.07546}

\bibitem[{{Downes} \& {Solomon}(1998)}]{Downes:1998aa}
{Downes}, D., \& {Solomon}, P.~M. 1998, \apj, 507, 615, \dodoi{10.1086/306339}

\bibitem[{{Dutton}(2012)}]{Dutton:2012aa}
{Dutton}, A.~A. 2012, \mnras, 424, 3123,
  \dodoi{10.1111/j.1365-2966.2012.21469.x}

\bibitem[{{Dutton} {et~al.}(2011){Dutton}, {van den Bosch}, {Faber}, {Simard},
  {Kassin}, {Koo}, {Bundy}, {Huang}, {Weiner}, {Cooper}, {Newman}, {Mozena}, \&
  {Koekemoer}}]{Dutton:2011aa}
{Dutton}, A.~A., {van den Bosch}, F.~C., {Faber}, S.~M., {et~al.} 2011, \mnras,
  410, 1660, \dodoi{10.1111/j.1365-2966.2010.17555.x}

\bibitem[{{Emonts} {et~al.}(2018){Emonts}, {Carilli}, {Narayanan}, {Lehnert},
  \& {Nyland}}]{Emonts:2018aa}
{Emonts}, B., {Carilli}, C., {Narayanan}, D., {Lehnert}, M., \& {Nyland}, K.
  2018, ArXiv e-prints.
\newblock \doarXiv{1810.06770}

\bibitem[{Engel {et~al.}(2010)Engel, Tacconi, Davies, Neri, Smail, Chapman,
  Genzel, Cox, Greve, Ivison, Blain, Bertoldi, \& Omont}]{Engel10}
Engel, H., Tacconi, L.~J., Davies, R.~I., {et~al.} 2010, \apj, 724, 233

\bibitem[{{Fern{\'a}ndez} {et~al.}(2016){Fern{\'a}ndez}, {Gim}, {van Gorkom},
  {Yun}, {Momjian}, {Popping}, {Chomiuk}, {Hess}, {Hunt}, {Kreckel}, {Lucero},
  {Maddox}, {Oosterloo}, {Pisano}, {Verheijen}, {Hales}, {Chung}, {Dodson},
  {Golap}, {Gross}, {Henning}, {Hibbard}, {Jaff{\'e}}, {Donovan Meyer},
  {Meyer}, {Sanchez-Barrantes}, {Schiminovich}, {Wicenec}, {Wilcots},
  {Bershady}, {Scoville}, {Strader}, {Tremou}, {Salinas}, \&
  {Ch{\'a}vez}}]{Fernandez:2016aa}
{Fern{\'a}ndez}, X., {Gim}, H.~B., {van Gorkom}, J.~H., {et~al.} 2016, \apjl,
  824, L1, \dodoi{10.3847/2041-8205/824/1/L1}

\bibitem[{{Foreman-Mackey} {et~al.}(2013){Foreman-Mackey}, {Hogg}, {Lang}, \&
  {Goodman}}]{Foreman-Mackey:2013aa}
{Foreman-Mackey}, D., {Hogg}, D.~W., {Lang}, D., \& {Goodman}, J. 2013, \pasp,
  125, 306, \dodoi{10.1086/670067}

\bibitem[{Fu {et~al.}(2012)Fu, Jullo, Cooray, Bussmann, Ivison,
  P{\'e}rez-Fournon, Djorgovski, Scoville, Yan, Riechers, Aguirre, Auld, Baes,
  Baker, Bradford, Cava, Clements, Dannerbauer, Dariush, De~Zotti, Dole, Dunne,
  Dye, Eales, Frayer, Gavazzi, Gurwell, Harris, Herranz, Hopwood, Hoyos, Ibar,
  Jarvis, Kim, Leeuw, Lupu, Maddox, Mart{\'\i}nez-Navajas, Micha{\l}owski,
  Negrello, Omont, Rosenman, Scott, Serjeant, Smail, Swinbank, Valiante, Verma,
  Vieira, Wardlow, \& van~der Werf}]{Fu12b}
Fu, H., Jullo, E., Cooray, A., {et~al.} 2012, \apj, 753, 134

\bibitem[{Fu {et~al.}(2013)Fu, Cooray, Feruglio, Ivison, Riechers, Gurwell,
  Bussmann, Harris, Altieri, Aussel, Baker, Bock, Boylan-Kolchin, Bridge,
  Calanog, Casey, Cava, Chapman, Clements, Conley, Cox, Farrah, Frayer,
  Hopwood, Jia, Magdis, Marsden, Mart{\'\i}nez-Navajas, Negrello, Neri, Oliver,
  Omont, Page, P{\'e}rez-Fournon, Schulz, Scott, Smith, Vaccari, Valtchanov,
  Vieira, Viero, Wang, Wardlow, \& Zemcov}]{Fu13}
Fu, H., Cooray, A., Feruglio, C., {et~al.} 2013, Nature, 498, 338

\bibitem[{{Greve} {et~al.}(2014){Greve}, {Leonidaki}, {Xilouris}, {Wei{\ss}},
  {Zhang}, {van der Werf}, {Aalto}, {Armus}, {D{\'{\i}}az-Santos}, {Evans},
  {Fischer}, {Gao}, {Gonz{\'a}lez-Alfonso}, {Harris}, {Henkel}, {Meijerink},
  {Naylor}, {Smith}, {Spaans}, {Stacey}, {Veilleux}, \&
  {Walter}}]{Greve:2014aa}
{Greve}, T.~R., {Leonidaki}, I., {Xilouris}, E.~M., {et~al.} 2014, \apj, 794,
  142, \dodoi{10.1088/0004-637X/794/2/142}

\bibitem[{Hainline {et~al.}(2011)Hainline, Blain, Smail, Alexander, Armus,
  Chapman, \& Ivison}]{Hainline11}
Hainline, L.~J., Blain, A.~W., Smail, I., {et~al.} 2011, \apj, 740, 96

\bibitem[{Harris {et~al.}(2012)Harris, Baker, Frayer, Smail, Swinbank,
  Riechers, van~der Werf, Auld, Baes, Bussmann, Buttiglione, Cava, Clements,
  Cooray, Dannerbauer, Dariush, De~Zotti, Dunne, Dye, Eales, Fritz,
  Gonz{\'a}lez-Nuevo, Hopwood, Ibar, Ivison, Jarvis, Maddox, Negrello, Rigby,
  Smith, Temi, \& Wardlow}]{Harris12}
Harris, A.~I., Baker, A.~J., Frayer, D.~T., {et~al.} 2012, \apj, 752, 152

\bibitem[{{Haynes} {et~al.}(2018){Haynes}, {Giovanelli}, {Kent}, {Adams},
  {Balonek}, {Craig}, {Fertig}, {Finn}, {Giovanardi}, {Hallenbeck}, {Hess},
  {Hoffman}, {Huang}, {Jones}, {Koopmann}, {Kornreich}, {Leisman}, {Miller},
  {Moorman}, {O'Connor}, {O'Donoghue}, {Papastergis}, {Troischt}, {Stark}, \&
  {Xiao}}]{Haynes18}
{Haynes}, M.~P., {Giovanelli}, R., {Kent}, B.~R., {et~al.} 2018, ArXiv
  e-prints.
\newblock \doarXiv{1805.11499}

\bibitem[{Ho(2007)}]{Ho07b}
Ho, L.~C. 2007, \apj, 669, 821

\bibitem[{{Hodge} {et~al.}(2012){Hodge}, {Carilli}, {Walter}, {de Blok},
  {Riechers}, {Daddi}, \& {Lentati}}]{Hodge12}
{Hodge}, J.~A., {Carilli}, C.~L., {Walter}, F., {et~al.} 2012, \apj, 760, 11,
  \dodoi{10.1088/0004-637X/760/1/11}

\bibitem[{{Ivison} {et~al.}(2011){Ivison}, {Papadopoulos}, {Smail}, {Greve},
  {Thomson}, {Xilouris}, \& {Chapman}}]{Ivison:2011aa}
{Ivison}, R.~J., {Papadopoulos}, P.~P., {Smail}, I., {et~al.} 2011, \mnras,
  412, 1913, \dodoi{10.1111/j.1365-2966.2010.18028.x}

\bibitem[{Kennicutt \& Evans(2012)}]{Kennicutt12}
Kennicutt, R.~C., \& Evans, N.~J. 2012, \araa, 50, 531

\bibitem[{{Lagos} {et~al.}(2011){Lagos}, {Baugh}, {Lacey}, {Benson}, {Kim}, \&
  {Power}}]{Lagos:2011ab}
{Lagos}, C.~D.~P., {Baugh}, C.~M., {Lacey}, C.~G., {et~al.} 2011, \mnras, 418,
  1649, \dodoi{10.1111/j.1365-2966.2011.19583.x}

\bibitem[{Magdis {et~al.}(2011)Magdis, Daddi, Elbaz, Sargent, Dickinson,
  Dannerbauer, Aussel, Walter, Hwang, Charmandaris, Hodge, Riechers,
  Rigopoulou, Carilli, Pannella, Mullaney, Leiton, \& Scott}]{Magdis11}
Magdis, G.~E., Daddi, E., Elbaz, D., {et~al.} 2011, \apj, 740, L15

\bibitem[{{Magdis} {et~al.}(2014){Magdis}, {Rigopoulou}, {Hopwood}, {Huang},
  {Farrah}, {Pearson}, {Alonso-Herrero}, {Bock}, {Clements}, {Cooray},
  {Griffin}, {Oliver}, {Perez Fournon}, {Riechers}, {Swinyard}, {Scott},
  {Thatte}, {Valtchanov}, \& {Vaccari}}]{Magdis14}
{Magdis}, G.~E., {Rigopoulou}, D., {Hopwood}, R., {et~al.} 2014, \apj, 796, 63,
  \dodoi{10.1088/0004-637X/796/1/63}

\bibitem[{Magnelli {et~al.}(2012)Magnelli, Saintonge, Lutz, Tacconi, Berta,
  Bournaud, Charmandaris, Dannerbauer, Elbaz, F{\"o}rster-Schreiber,
  Graci{\'a}-Carpio, Ivison, Maiolino, Nordon, Popesso, Rodighiero, Santini, \&
  Wuyts}]{Magnelli12b}
Magnelli, B., Saintonge, A., Lutz, D., {et~al.} 2012, \aap, 548, 22

\bibitem[{{McGaugh}(2012)}]{McGaugh:2012aa}
{McGaugh}, S.~S. 2012, \aj, 143, 40, \dodoi{10.1088/0004-6256/143/2/40}

\bibitem[{{McGaugh} \& {Schombert}(2015)}]{McGaugh:2015aa}
{McGaugh}, S.~S., \& {Schombert}, J.~M. 2015, \apj, 802, 18,
  \dodoi{10.1088/0004-637X/802/1/18}

\bibitem[{{McKee} \& {Ostriker}(2007)}]{McKee07}
{McKee}, C.~F., \& {Ostriker}, E.~C. 2007, \araa, 45, 565,
  \dodoi{10.1146/annurev.astro.45.051806.110602}

\bibitem[{{Micha{\l}owski} {et~al.}(2010){Micha{\l}owski}, {Hjorth}, \&
  {Watson}}]{Michaowski10}
{Micha{\l}owski}, M., {Hjorth}, J., \& {Watson}, D. 2010, \aap, 514, A67,
  \dodoi{10.1051/0004-6361/200913634}

\bibitem[{Micha{\l}owski {et~al.}(2012)Micha{\l}owski, Dunlop, Cirasuolo,
  Hjorth, Hayward, \& Watson}]{Michalowski12}
Micha{\l}owski, M.~J., Dunlop, J.~S., Cirasuolo, M., {et~al.} 2012, \aap, 541,
  85

\bibitem[{{Miller} {et~al.}(2011){Miller}, {Bundy}, {Sullivan}, {Ellis}, \&
  {Treu}}]{Miller:2011aa}
{Miller}, S.~H., {Bundy}, K., {Sullivan}, M., {Ellis}, R.~S., \& {Treu}, T.
  2011, \apj, 741, 115, \dodoi{10.1088/0004-637X/741/2/115}

\bibitem[{{Miller} {et~al.}(2012){Miller}, {Ellis}, {Sullivan}, {Bundy},
  {Newman}, \& {Treu}}]{Miller:2012aa}
{Miller}, S.~H., {Ellis}, R.~S., {Sullivan}, M., {et~al.} 2012, \apj, 753, 74,
  \dodoi{10.1088/0004-637X/753/1/74}

\bibitem[{Noeske {et~al.}(2007)Noeske, Weiner, Faber, Papovich, Koo,
  Somerville, Bundy, Conselice, Newman, Schiminovich, Le~Floc'h, Coil, Rieke,
  Lotz, Primack, Barmby, Cooper, Davis, Ellis, Fazio, Guhathakurta, Huang,
  Kassin, Martin, Phillips, Rich, Small, Willmer, \& Wilson}]{Noeske07a}
Noeske, K.~G., Weiner, B.~J., Faber, S.~M., {et~al.} 2007, \apj, 660, L43

\bibitem[{{Obreschkow} \& {Meyer}(2013)}]{Obreschkow:2013aa}
{Obreschkow}, D., \& {Meyer}, M. 2013, \apj, 777, 140,
  \dodoi{10.1088/0004-637X/777/2/140}

\bibitem[{{Obreschkow} \& {Rawlings}(2009)}]{Obreschkow:2009aa}
{Obreschkow}, D., \& {Rawlings}, S. 2009, \apjl, 696, L129,
  \dodoi{10.1088/0004-637X/696/2/L129}

\bibitem[{{Papadopoulos} {et~al.}(2012){Papadopoulos}, {van der Werf},
  {Xilouris}, {Isaak}, \& {Gao}}]{Papadopoulos:2012ab}
{Papadopoulos}, P.~P., {van der Werf}, P., {Xilouris}, E., {Isaak}, K.~G., \&
  {Gao}, Y. 2012, \apj, 751, 10, \dodoi{10.1088/0004-637X/751/1/10}

\bibitem[{{Papastergis} {et~al.}(2016){Papastergis}, {Adams}, \& {van der
  Hulst}}]{Papastergis:2016aa}
{Papastergis}, E., {Adams}, E.~A.~K., \& {van der Hulst}, J.~M. 2016, \aap,
  593, A39, \dodoi{10.1051/0004-6361/201628410}
  
\bibitem[{{Pereira-Santaella} {et~al.}(2018){Pereira-Santaella}, {Colina}, {Garc\`ia-Burillo},
  {Combes}, {Emonts}, {Aalto}, {Alonso-Herrero}, {Arribas}, {Henkel}, {Labiano}, {Muller},
  {Piqueras L{\`o}pez}, {Rigopoulou},\&{van der Werf}}]{Pereira-Santaella2018:aa}
  {Pereira-Santaella}, M., {Colina}, L., {Garc{\`i}a-Burillo}, S., {et~al.} 2018, \aap, 616, A171,
  \dodoi{10.1051/0004-6361/201833089}

\bibitem[{{Saintonge} {et~al.}(2011){Saintonge}, {Kauffmann}, {Kramer},
  {Tacconi}, {Buchbender}, {Catinella}, {Fabello}, {Graci{\'a}-Carpio}, {Wang},
  {Cortese}, {Fu}, {Genzel}, {Giovanelli}, {Guo}, {Haynes}, {Heckman},
  {Krumholz}, {Lemonias}, {Li}, {Moran}, {Rodriguez-Fernandez}, {Schiminovich},
  {Schuster}, \& {Sievers}}]{Saintonge11}
{Saintonge}, A., {Kauffmann}, G., {Kramer}, C., {et~al.} 2011, \mnras, 415, 32,
  \dodoi{10.1111/j.1365-2966.2011.18677.x}

\bibitem[{{Saintonge} {et~al.}(2017){Saintonge}, {Catinella}, {Tacconi},
  {Kauffmann}, {Genzel}, {Cortese}, {Dav{\'e}}, {Fletcher},
  {Graci{\'a}-Carpio}, {Kramer}, {Heckman}, {Janowiecki}, {Lutz}, {Rosario},
  {Schiminovich}, {Schuster}, {Wang}, {Wuyts}, {Borthakur}, {Lamperti}, \&
  {Roberts-Borsani}}]{Saintonge17}
{Saintonge}, A., {Catinella}, B., {Tacconi}, L.~J., {et~al.} 2017, \apjs, 233,
  22, \dodoi{10.3847/1538-4365/aa97e0}

\bibitem[{{Saito} {et~al.}(2017){Saito}, {Iono}, {Xu}, {Sliwa}, {Ueda},
  {Espada}, {Kaneko}, {K{\"o}nig}, {Nakanishi}, {Lee}, {Yun}, {Aalto},
  {Hibbard}, {Yamashita}, {Motohara}, \& {Kawabe}}]{Saito:2017aa}
{Saito}, T., {Iono}, D., {Xu}, C.~K., {et~al.} 2017, \apj, 835, 174,
  \dodoi{10.3847/1538-4357/835/2/174}

\bibitem[{{Schoniger} \& {Sofue}(1994)}]{Schoniger94}
{Schoniger}, F., \& {Sofue}, Y. 1994, \aap, 283, 21

\bibitem[{Solomon {et~al.}(1997)Solomon, Downes, Radford, \&
  Barrett}]{Solomon97}
Solomon, P.~M., Downes, D., Radford, S. J.~E., \& Barrett, J.~W. 1997, \apj,
  478, 144

\bibitem[{{Sorce} {et~al.}(2013){Sorce}, {Courtois}, {Tully}, {Seibert},
  {Scowcroft}, {Freedman}, {Madore}, {Persson}, {Monson}, \& {Rigby}}]{Sorce13}
{Sorce}, J.~G., {Courtois}, H.~M., {Tully}, R.~B., {et~al.} 2013, \apj, 765,
  94, \dodoi{10.1088/0004-637X/765/2/94}

\bibitem[{{Springob} {et~al.}(2005){Springob}, {Haynes}, {Giovanelli}, \&
  {Kent}}]{Springob:2005aa}
{Springob}, C.~M., {Haynes}, M.~P., {Giovanelli}, R., \& {Kent}, B.~R. 2005,
  \apjs, 160, 149, \dodoi{10.1086/431550}

\bibitem[{Tacconi {et~al.}(2010)Tacconi, Genzel, Neri, Cox, Cooper, Shapiro,
  Bolatto, Bouch{\'e}, Bournaud, Burkert, Combes, Comerford, Davis, Schreiber,
  Garcia-Burillo, Gracia-Carpio, Lutz, Naab, Omont, Shapley, Sternberg, \&
  Weiner}]{Tacconi10}
Tacconi, L.~J., Genzel, R., Neri, R., {et~al.} 2010, Nature, 463, 781

\bibitem[{{Tacconi} {et~al.}(2013){Tacconi}, {Neri}, {Genzel}, {Combes},
  {Bolatto}, {Cooper}, {Wuyts}, {Bournaud}, {Burkert}, {Comerford}, {Cox},
  {Davis}, {F{\"o}rster Schreiber}, {Garc{\'{\i}}a-Burillo}, {Gracia-Carpio},
  {Lutz}, {Naab}, {Newman}, {Omont}, {Saintonge}, {Shapiro Griffin}, {Shapley},
  {Sternberg}, \& {Weiner}}]{Tacconi13}
{Tacconi}, L.~J., {Neri}, R., {Genzel}, R., {et~al.} 2013, \apj, 768, 74,
  \dodoi{10.1088/0004-637X/768/1/74}

\bibitem[{{Tacconi} {et~al.}(2018){Tacconi}, {Genzel}, {Saintonge}, {Combes},
  {Garc{\'{\i}}a-Burillo}, {Neri}, {Bolatto}, {Contini}, {F{\"o}rster
  Schreiber}, {Lilly}, {Lutz}, {Wuyts}, {Accurso}, {Boissier}, {Boone},
  {Bouch{\'e}}, {Bournaud}, {Burkert}, {Carollo}, {Cooper}, {Cox}, {Feruglio},
  {Freundlich}, {Herrera-Camus}, {Juneau}, {Lippa}, {Naab}, {Renzini},
  {Salome}, {Sternberg}, {Tadaki}, {{\"U}bler}, {Walter}, {Weiner}, \&
  {Weiss}}]{Tacconi18}
{Tacconi}, L.~J., {Genzel}, R., {Saintonge}, A., {et~al.} 2018, \apj, 853, 179,
  \dodoi{10.3847/1538-4357/aaa4b4}

\bibitem[{{Tiley} {et~al.}(2016){Tiley}, {Bureau}, {Saintonge}, {Topal},
  {Davis}, \& {Torii}}]{Tiley:2016aa}
{Tiley}, A.~L., {Bureau}, M., {Saintonge}, A., {et~al.} 2016, \mnras, 461,
  3494, \dodoi{10.1093/mnras/stw1545}

\bibitem[{{Topal} {et~al.}(2018){Topal}, {Bureau}, {Tiley}, {Davis}, \&
  {Torii}}]{Topal18}
{Topal}, S., {Bureau}, M., {Tiley}, A.~L., {Davis}, T.~A., \& {Torii}, K. 2018,
  \mnras, 479, 3319, \dodoi{10.1093/mnras/sty1617}

\bibitem[{{Tully} \& {Courtois}(2012)}]{Tully12}
{Tully}, R.~B., \& {Courtois}, H.~M. 2012, \apj, 749, 78,
  \dodoi{10.1088/0004-637X/749/1/78}

\bibitem[{{Tully} \& {Fisher}(1977)}]{Tully77}
{Tully}, R.~B., \& {Fisher}, J.~R. 1977, \aap, 54, 661

\bibitem[{{Turner} {et~al.}(2017){Turner}, {Harrison}, {Cirasuolo}, {McLure},
  {Dunlop}, {Swinbank}, \& {Tiley}}]{Turner17}
{Turner}, O.~J., {Harrison}, C.~M., {Cirasuolo}, M., {et~al.} 2017, ArXiv
  e-prints.
\newblock \doarXiv{1711.03604}

\bibitem[{{{\"U}bler} {et~al.}(2017){{\"U}bler}, {F{\"o}rster Schreiber},
  {Genzel}, {Wisnioski}, {Wuyts}, {Lang}, {Naab}, {Burkert}, {van Dokkum},
  {Tacconi}, {Wilman}, {Fossati}, {Mendel}, {Beifiori}, {Belli}, {Bender},
  {Brammer}, {Chan}, {Davies}, {Fabricius}, {Galametz}, {Lutz}, {Momcheva},
  {Nelson}, {Saglia}, {Seitz}, \& {Tadaki}}]{Ubler17}
{{\"U}bler}, H., {F{\"o}rster Schreiber}, N.~M., {Genzel}, R., {et~al.} 2017,
  \apj, 842, 121, \dodoi{10.3847/1538-4357/aa7558}

\bibitem[{{Villanueva} {et~al.}(2017){Villanueva}, {Ibar}, {Hughes},
  {Lara-L{\'o}pez}, {Dunne}, {Eales}, {Ivison}, {Aravena}, {Baes}, {Bourne},
  {Cassata}, {Cooray}, {Dannerbauer}, {Davies}, {Driver}, {Dye}, {Furlanetto},
  {Herrera-Camus}, {Maddox}, {Micha{\l}owski}, {Molina}, {Riechers}, {Sansom},
  {Smith}, {Rodighiero}, {Valiante}, \& {van der Werf}}]{Villanueva17}
{Villanueva}, V., {Ibar}, E., {Hughes}, T.~M., {et~al.} 2017, \mnras, 470,
  3775, \dodoi{10.1093/mnras/stx1338}

\bibitem[{{Walter} {et~al.}(2016){Walter}, {Decarli}, {Aravena}, {Carilli},
  {Bouwens}, {da Cunha}, {Daddi}, {Ivison}, {Riechers}, {Smail}, {Swinbank},
  {Weiss}, {Anguita}, {Assef}, {Bacon}, {Bauer}, {Bell}, {Bertoldi}, {Chapman},
  {Colina}, {Cortes}, {Cox}, {Dickinson}, {Elbaz}, {G{\'o}nzalez-L{\'o}pez},
  {Ibar}, {Inami}, {Infante}, {Hodge}, {Karim}, {Le Fevre}, {Magnelli}, {Neri},
  {Oesch}, {Ota}, {Popping}, {Rix}, {Sargent}, {Sheth}, {van der Wel}, {van der
  Werf}, \& {Wagg}}]{Walter16}
{Walter}, F., {Decarli}, R., {Aravena}, M., {et~al.} 2016, \apj, 833, 67,
  \dodoi{10.3847/1538-4357/833/1/67}

\bibitem[{{Wong} \& {Blitz}(2002)}]{Wong:2002aa}
{Wong}, T., \& {Blitz}, L. 2002, \apj, 569, 157, \dodoi{10.1086/339287}

\bibitem[{{Xue} {et~al.}(2018){Xue}, {Fu}, {Isbell}, {Ivison}, {Cooray}, \&
  {Oteo}}]{Xue18}
{Xue}, R., {Fu}, H., {Isbell}, J., {et~al.} 2018, \apjl, 864, L11,
  \dodoi{10.3847/2041-8213/aad9a9}

\bibitem[{{Yamashita} {et~al.}(2017){Yamashita}, {Komugi}, {Matsuhara},
  {Armus}, {Inami}, {Ueda}, {Iono}, {Kohno}, {Evans}, \&
  {Arimatsu}}]{Yamashita17}
{Yamashita}, T., {Komugi}, S., {Matsuhara}, H., {et~al.} 2017, \apj, 844, 96,
  \dodoi{10.3847/1538-4357/aa7af1}

\bibitem[{{Zaritsky} {et~al.}(2014){Zaritsky}, {Courtois}, {Mu{\~n}oz-Mateos},
  {Sorce}, {Erroz-Ferrer}, {Comer{\'o}n}, {Gadotti}, {Gil de Paz}, {Hinz},
  {Laurikainen}, {Kim}, {Laine}, {Men{\'e}ndez-Delmestre}, {Mizusawa}, {Regan},
  {Salo}, {Seibert}, {Sheth}, {Athanassoula}, {Bosma}, {Cisternas}, {Ho}, \&
  {Holwerda}}]{Zaritsky14a}
{Zaritsky}, D., {Courtois}, H., {Mu{\~n}oz-Mateos}, J.-C., {et~al.} 2014, \aj,
  147, 134, \dodoi{10.1088/0004-6256/147/6/134}

\end{thebibliography}

\end{document}